\documentclass[prl,aps,superscriptaddress,groupedaddress,twocolumn,nofootinbib,showpacs]{revtex4}
\usepackage{xspace,amsmath,amsfonts,amsthm,amssymb,amsbsy}
\usepackage{graphicx}
\usepackage{amssymb}
\usepackage{color}
\usepackage{psfrag}
\usepackage{ifsym}
\usepackage{hyperref}
\usepackage{epstopdf}
\usepackage{units}
\usepackage{tikz}
\usepackage{tikz-3dplot}

\definecolor{pink}{rgb}{1,0.078,0.57}

\definecolor{green}{rgb}{0,0.7,0.9}

\newcommand{\aurelia}{\color{black}}

\begin{document}

\title{Thermal light cannot be represented as a statistical mixture of {\aurelia single} pulses%
}
\author{Aur\'elia Chenu}
\affiliation{Department of Chemistry, Centre for Quantum Information and Quantum Control,
80 St George Street, University of Toronto, Toronto, Ontario, M5S 3H6 Canada}
\email[ Corresponding author: ]{aurelia.chenu@utoronto.ca}
\author{Agata M. Bra\'nczyk}
\affiliation{Department of Chemistry, Centre for Quantum Information and Quantum Control,
80 St George Street, University of Toronto, Toronto, Ontario, M5S 3H6 Canada}
\affiliation{Perimeter Institute for Theoretical Physics, Waterloo, Ontario, N2L 2Y5,
Canada}
\author{Gregory D. Scholes}
\affiliation{Department of Chemistry, Centre for Quantum Information and Quantum Control,
80 St George Street, University of Toronto, Toronto, Ontario, M5S 3H6 Canada}
\affiliation{Department of Chemistry, Princeton University, Washington Rd, Princeton, New
Jersey 08544, U.S.A.}
\author{J. E. Sipe}
\affiliation{Department of Physics, 60 St George Street, University of Toronto, Toronto,
Ontario, M5R 3C3 Canada}

\begin{abstract}
We ask whether or not thermal light can be represented as a mixture of
single broadband coherent pulses. We find that it \emph{cannot}. 
Such a mixture is simply not rich enough to mimic thermal light; indeed,
it cannot even reproduce the first-order correlation function. We show that
it is possible to construct a modified mixture of single coherent pulses
that \textit{does} yield the correct first-order correlation function at
equal space points. However, as we then demonstrate, such a mixture cannot
reproduce the second-order correlation function.
\end{abstract}

\pacs{44.40.+a, 42.50.Ar,  42.50.-p}
\maketitle



Absorption of light by molecules can initiate fundamental photo-induced
processes including photochemical reactions, photocatalysis, and solar
energy conversion. While the photo-initiated dynamics can be resolved by
using short laser pulses to populate and probe excited-state populations 
\cite{Fleming1986a, Zewail2000a,Grondelle2006a, Cheng2009a}, the use
of these techniques raises the question of whether or not the ultrafast
pulses employed -- which are significantly different from natural thermal
light -- lead to behaviour specific to those ultrafast pulses. In
particular, some researchers have recently questioned whether dynamics
initiated by sunlight excitation might be different from those detected in
femtosecond laser experiments performed on light-harvesting complexes \cite%
{Jiang1991a, Mancal2010a, Hoki2011a, Brumer2012a, Kassal2013a}%
. This has opened a debate on how photo-excitation by natural light should
be understood. For example, can sunlight be viewed as \textquotedblleft a
series of random ultrashort spikes with a duration as short as the bandwidth
allows\textquotedblright\ \cite{Cheng2009a}? 
 {\aurelia Our work is inspired by femtosecond laser experiments, but rather than considering the relevance of these experiments to natural-light excitation, we focus on the relationship between sunlight and laser light.}

Light from the sun indeed has an ultra-short coherence time of approximately 
$1.3$~fs \cite{Kano1962a}; it has a spectrum close to that of black-body
radiation at approximately 5777~K \cite{Iqbal1983a}, characterized by
thermal photon-number statistics \cite{MandelWolfBook}. \ 
In this paper we ask whether or not thermal light can be understood as a mixture of single
broadband coherent pulses. \ 

We find that it cannot. \ 
{\aurelia The first clue is given by considering a widely used class of pulses, 
where each pulse is defined by a linear phase relationship between its composing modes.} 
The density matrix of a mixture of such pulses cannot represent thermal
equilibrium, for any mixture of these  {\aurelia individual} pulses would exhibit off-diagonal
elements in the density matrix when written in a spectral-Fock basis,
whereas the density matrix representing thermal equilibrium is diagonal; see
the Supplemental Material.

{\aurelia 
More generally, while the state of thermal light 
 is represented by a density operator $\rho ^{\mathrm{th}}$ with unit trace, we
demonstrate that \textit{no }unit-trace density operator $\rho^{\mathrm{mix}%
} $ consisting of a mixture of single pulses can equal $\rho ^{\mathrm{th}}$. \
Such a mixture cannot even give the correct result for the
first-order correlation function. 
Nonetheless, it is possible to construct a  \emph{trace-improper} mixture $\rho ^{\mathrm{imp}}$
that \emph{does} yield a first-order correlation function at equal space
points that matches that of thermal light. 
 \ This has apparently not been
demonstrated yet; we do it here.\ }

To begin we build our pulses by quantizing the electromagnetic field in an
infinite volume, with annihilation (creation) operators ${a}_{\mathbf{k}%
\lambda }$ (${a}_{\mathbf{k\lambda }}^{\dagger }$), where the wave vector $%
\mathbf{k}$ ranges continuously and the helicity $\lambda $ is positive or
negative; these operators satisfy the commutation relations $\left[ {a}_{%
\mathbf{k}\lambda },{a}_{\mathbf{k}^{\prime }\lambda ^{\prime }}^{\dagger }%
\right] =\delta (\mathbf{k-k}^{\prime })\delta _{\lambda \lambda ^{\prime }}$%
. \ A pulse is characterized by its nominal position $\mathbf{r}_{\mathrm{o}}
$, a (complex) amplitude $\alpha _{\mathbf{r}_{\mathrm{o}}s}$, a spectral
distribution $f_{\mathbf{r}_{\mathrm{o}}s;\mathbf{k\lambda }}$ -- normalized
so that $\sum_{\lambda }\int d\mathbf{k}\;\left\vert f_{\mathbf{r}_{\mathrm{o%
}}s;\mathbf{k}\lambda }\right\vert ^{2}=1$, with $d\mathbf{k}%
=dk_{x}dk_{y}dk_{z}$ -- and other parameters that we label collectively by $s
$. From the spectral distribution we construct a creation operator 
\begin{equation*}
{a}_{\mathbf{r}_{\mathrm{o}}s}^{\dagger }=\sum_{\lambda }\int d\mathbf{k}%
\;f_{\mathbf{r}_{\mathrm{o}}s;\mathbf{k}\lambda }{a}_{\mathbf{k}\lambda
}^{\dagger },
\end{equation*}%
with $\left[ {a}_{\mathbf{r}_{\mathrm{o}}s},{a}_{\mathbf{r}_{\mathrm{o}%
}s}^{\dagger }\right] =1$, and the pulse is described by the quantum state 
\begin{equation*}
\left\vert \alpha _{\mathbf{r}_{\mathrm{o}}s}f_{\mathbf{r}_{\mathrm{o}%
}s}\right\rangle \equiv e^{\alpha _{\mathbf{r}_{\mathrm{o}}s}{a}_{\mathbf{r}%
_{\mathrm{o}}s}^{\dagger }-\alpha _{\mathbf{r}_{\mathrm{o}}s}^{\ast }{a}_{%
\mathbf{r}_{\mathrm{o}}s}}\left\vert vac\right\rangle ,
\end{equation*}%
where $\left\vert vac\right\rangle $ is the vacuum state and $\alpha _{%
\mathbf{r}_{\mathrm{o}}s}$ is a complex number; $\left\langle \alpha _{%
\mathbf{r}_{\mathrm{o}}s}f_{\mathbf{r}_{\mathrm{o}}s}|\alpha _{\mathbf{r}_{%
\mathrm{o}}s}f_{\mathbf{r}_{\mathrm{o}}s}\right\rangle =1$. \ For the
positive frequency part of the (Heisenberg) electric field operator, 
\begin{equation}
{\mathbf{E}}^{(+)}(\mathbf{r},t)=i\sum_{\lambda }\int d\mathbf{k}\;\sqrt{%
\frac{\hbar \omega _{k}}{16\pi ^{3}\epsilon _{0}}}\mathbf{e}_{\mathbf{k}%
\lambda }e^{i\mathbf{k\cdot r}}e^{-i\omega _{k}t}{a}_{\mathbf{k}\lambda },
\label{fieldpositive}
\end{equation}%
where $\omega _{k}=c\left\vert \mathbf{k}\right\vert $ and $\mathbf{e}_{%
\mathbf{k\lambda }}$ are the polarization vectors, we have the expectation
value 
\begin{equation*}
\mathcal{E}^{(\mathbf{r}_{\mathrm{o}}s)}(\mathbf{r},t)\equiv \langle \alpha
_{\mathbf{r}_{\mathrm{o}}s}f_{\mathbf{r}_{\mathrm{o}}s}|_{{}}\mathbf{E}%
^{(+)}(\mathbf{r},t)|\alpha _{\mathbf{r}_{\mathrm{o}}s}f_{\mathbf{r}_{%
\mathrm{o}}s}\rangle _{{}}
\end{equation*}%
given by (\ref{fieldpositive}) with the operator ${a}_{\mathbf{k\lambda }}$
replaced by the complex number $\alpha _{\mathbf{r}_{\mathrm{o}}s}f_{\mathbf{%
r}_{\mathrm{o}}s;\mathbf{k\lambda }}$. The state $\left\vert \alpha _{%
\mathbf{r}_{\mathrm{o}}s}f_{\mathbf{r}_{\mathrm{o}}s}\right\rangle $ is
\textquotedblleft coherent" in the sense that it factorizes correlation
functions according to ${G}^{(n)(\mathbf{r}_{\mathrm{o}}s)}(\mathbf{r}_{1}{t}%
_{1}\dots \mathbf{r}_{n}{t}_{n};\mathbf{r}_{{n}+1}{t}_{{n}+1}\dots \mathbf{r}%
_{2{n}}{t}_{2{n}})=\prod_{j}\left( \mathcal{E}^{(\mathbf{r}_{\mathrm{o}}s)}(%
\mathbf{r}_{j},{t}_{j})\right) ^{\ast }\mathcal{E}^{(\mathbf{r}_{\mathrm{o}%
}s)}(\mathbf{r}_{j+n},{t}_{j+n})$ for all orders of ${n}$ \cite{Glauber1963a}%
; here we use the superscript $(\mathbf{r}_{\mathrm{o}}s)$ on $G$ to
identify the state $|\alpha _{\mathbf{r}_{\mathrm{o}}s}f_{\mathbf{r}_{%
\mathrm{o}}s}\rangle _{{}}$, which is the quantum description of what might
be called a \textquotedblleft classical" pulse \cite{Glauber1963a}. 
In
particular, for such a state we have 
\begin{align}
G_{ij}^{(1)(\mathbf{r}_{\mathrm{o}}s)}(\mathbf{r}t;\mathbf{r^{\prime }}%
t^{\prime }){}\equiv {}& \langle \alpha _{\mathbf{r}_{\mathrm{o}}s}f_{%
\mathbf{r}_{\mathrm{o}}s}|_{{}}{E}_{i}^{(-)}(\mathbf{r},t){E}_{j}^{(+)}(%
\mathbf{r}^{\prime },t^{\prime })|\alpha _{\mathbf{r}_{\mathrm{o}}s}f_{%
\mathbf{r}_{\mathrm{o}}s}\rangle _{{}}  \notag \\
={}& \left( \mathcal{E}_{i}^{(\mathbf{r}_{\mathrm{o}}s)}(\mathbf{r},{t}%
)\right) ^{\ast }\left( \mathcal{E}_{j}^{(\mathbf{r}_{\mathrm{o}}s)}(\mathbf{%
r}^{\prime },{t}^{\prime })\right) ,  \notag
\end{align}%
where subscripts on field labels indicate Cartesian components, e.g. $%
E_{j}^{(+)}(\mathbf{r},t)=\mathbf{E}^{(+)}(\mathbf{r},t)\cdot 
\mathbf{\hat{\jmath}}$. \ 

We consider families of pulses such that for fixed parameters $s$ the pulses
only differ by their nominal positions $\mathbf{r}_{\mathrm{o}}$. For such
families of pulses we have 
\begin{equation}
f_{\mathbf{r}_{\mathrm{o}}s;\mathbf{k\lambda }}=K(s,\mathbf{k}\lambda
)\,e^{-i\mathbf{k\cdot r}_{\mathrm{o}}},  \label{Kdef}
\end{equation}%
and the associated $\mathcal{E}^{(\mathbf{r}_{\mathrm{o}}s)}(\mathbf{r},{t})$
depends on $\mathbf{r}$ and $\mathbf{r}_{\mathrm{o}}$ only through its
dependence on $(\mathbf{r-r}_{\mathrm{o}})$; we will give particular
examples of $K(s,\mathbf{k\lambda )}$ below. \ We assume that each member of
the family is well localized in space at some initial time, $G_{ij}^{(1)(%
\mathbf{r}_{\mathrm{o}}s)}(\mathbf{r}0;\mathbf{r}0)\rightarrow 0$ as $%
\left\vert \mathbf{r-r}_{\mathrm{o}}\right\vert \rightarrow \infty $, and
that the integral of $G_{ij}^{(1)(\mathbf{r}_{\mathrm{o}}s)}(\mathbf{r}0;%
\mathbf{r}0)$ over all space is finite. \ Then for fixed $\mathbf{r}$ the
integral over all $\mathbf{r}_{\mathrm{o}}$ of $G_{ij}^{(1)(\mathbf{r}_{%
\mathrm{o}}s)}(\mathbf{r}0;\mathbf{r}0)$ will also be finite. \ It will be
convenient below to work with a volume $\Omega$
centered at the origin,{\aurelia \footnote{Note that $\Omega$ is different from the quantization volume that is always taken infinite in this Letter}}
 and
we define 
\begin{equation}
\mu _{i}(\mathbf{r,}s\mathbf{,}\Omega )\equiv \int_{\Omega }G_{ii}^{(1)(%
\mathbf{r}_{\mathrm{o}}s)}(\mathbf{r}0;\mathbf{r}0)\;d\mathbf{r}_{\mathrm{o}}.
\label{finitepulseintegral}
\end{equation}%
Since each $G_{ii}^{(1)(\mathbf{r}_{\mathrm{o}}s)}(\mathbf{r}0;\mathbf{r}0)$
is real and positive, $\mu _{i}(\mathbf{r},s,\Omega )$ will be finite and
positive for all $\mathbf{r}$ and $\Omega $, increasing as $\Omega $
increases and with a well-defined limit $\mu _{i}(\mathbf{r},s,\infty )$;
here and except when explicitly mentioned otherwise we assume the amplitudes 
$\alpha _{\mathbf{r}_{\mathrm{o}}s}$ and parameters $s$ used to characterize
the pulses are held fixed and independent of $\Omega $.

To attempt to describe thermal light in a volume $\Omega $ by a mixture of {\aurelia single} pulses, 
we would write 
\begin{equation}
\rho ^{\mathrm{mix}}=\int ds\int_{\Omega }d\mathbf{r}_{\mathrm{o}}\frac{%
\mathfrak{p}(s)}{\Omega }\left\vert \alpha _{\mathbf{r}_{\mathrm{o}}s}f_{%
\mathbf{r}_{\mathrm{o}}s}\right\rangle \left\langle \alpha _{\mathbf{r}_{%
\mathrm{o}}s}f_{\mathbf{r}_{\mathrm{o}}s}\right\vert ,  \label{rhomix}
\end{equation}%
where pulses are included with equal density about each central position $%
\mathbf{r}_{\mathrm{o}}$ in the volume, and $\mathfrak{p}(s)\geq 0$ with 
\begin{equation}
\int ds\;\mathfrak{p}(s)=1,  \label{normp}
\end{equation}%
where here and in (\ref{rhomix}) the variables constituting $s$ are to be
integrated or summed over as required; the condition (\ref{normp})
guarantees that \textrm{Tr}$(\rho ^{\mathrm{mix}})=1$. \ 
{\aurelia 
Because correlation functions describe the interaction of light with matter, 
 they form a practical tool 
to compare different radiation states \cite{Glauber1963a}.
}
For thermal light
filling all space  \cite{Kano1962a,MandelWolfBook} 
\begin{eqnarray}
G_{ij}^{(1)\mathrm{th}}(\mathbf{r}t;\mathbf{r}(t+\tau )) &=&\mathrm{Tr}%
\left( \rho ^{\mathrm{th}}{E}_{i}^{(-)}(\mathbf{r},t){E}_{j}^{(+)}(\mathbf{r}%
,t+\tau )\right)  \notag \\
&=&\delta _{ij}\int_{0}^{\infty }\frac{\hbar ck^{3}}{6\pi ^{2}\epsilon _{0}}%
\frac{e^{-ick\tau }}{e^{\beta \hbar ck}-1}dk,  \label{thermal1}
\end{eqnarray}%
and while near the edge of the volume $\Omega $ we would not expect (\ref%
{rhomix}) to give a correct representation of thermal equilibrium, we would
demand that it does so near the origin.

We now prove that we cannot choose the pulses so that $\rho ^{\mathrm{mix}%
}=\rho ^{\mathrm{th}}$ as $\Omega \rightarrow \infty $. \ For 
\begin{eqnarray*}
G_{ii}^{(1)\mathrm{mix}}(\mathbf{0}0;\mathbf{0}0) &=&\mathrm{Tr}\left( \rho
^{\mathrm{mix}}{E}_{i}^{(-)}(\mathbf{0},0){E}_{i}^{(+)}(\mathbf{0},0)\right)
\\
&=&\int ds\int_{\Omega }d\mathbf{r}_{\mathrm{o}}\frac{\mathfrak{p}(s)}{%
\Omega }G_{ii}^{(1)(\mathbf{r}_{\mathrm{o}}s)}(\mathbf{0}0;\mathbf{0}0) \\
&=&\frac{1}{\Omega }\int ds\;\mathfrak{p}(s)\mu _{i}(\mathbf{0,}s\mathbf{,}%
\Omega )\,,
\end{eqnarray*}%
and since $G_{ii}^{(1)\mathrm{mix}}(\mathbf{0}0;\mathbf{0}0)$ is clearly
finite for any $\Omega $, the integral of $\mathfrak{p}(s)\mu _{i}(\mathbf{0,%
}s\mathbf{,}\Omega )$ over $s$ must be finite for any $\Omega $. \ As $%
\mathfrak{p}(s)\mu _{i}(\mathbf{0,}s\mathbf{,}\Omega )>0$ for all $s$ and is
an increasing function of $\Omega $ with a well-defined limit $\mathfrak{p}%
(s)\mu _{i}(\mathbf{0,}s\mathbf{,}\infty )$ as $\Omega \rightarrow \infty $,
we see that $G_{ii}^{(1)\mathrm{mix}}(\mathbf{0}0;\mathbf{0}0)\rightarrow 0$
as $\Omega \rightarrow \infty $. But from (\ref{thermal1}) it is clear that $%
G_{ii}^{(1)\mathrm{th}}(\mathbf{0}0;\mathbf{0}0)\neq 0$. \ Thus we cannot
represent thermal equilibrium by a unit-trace density operator describing a
mixture of {\aurelia single} pulses. \ Such a mixture is simply not rich enough to describe
thermal light.

The proof would fail if we allowed the amplitudes of the pulses to change as 
$\Omega $ changed. In fact, we will see below that we \emph{can} mimic the
first-order correlation function of thermal light at equal-space points by
that of unit-trace mixture of {\aurelia single} pulses if we allow the square of the
amplitudes of the pulses to scale as $\Omega $. While such a scaling could
be entertained for finite $\Omega $,  the pulses would acquire infinite
energy as $%
\Omega \rightarrow \infty $.
 If we return to our assumption of fixed amplitudes and properties
regardless of $\Omega $, the scaling of $G_{ij}^{(1)\mathrm{mix}}(\mathbf{0}%
0;\mathbf{0}0)$ as $1/\Omega $ suggests a different strategy, i.e., the
consideration of \textit{trace-improper }density operators, of the form 
\begin{equation}
\rho ^{\mathrm{imp}}=\int ds\int_{\Omega }d\mathbf{r}_{\mathrm{o}}\;\bar{p}%
(s)\left\vert \alpha _{\mathbf{r}_{\mathrm{o}}s}f_{\mathbf{r}_{\mathrm{o}%
}s}\right\rangle \left\langle \alpha _{\mathbf{r}_{\mathrm{o}}s}f_{\mathbf{r}%
_{\mathrm{o}}s}\right\vert\,,  \label{rhoimp}
\end{equation}%
where $\bar{p}(s)\geq 0$ and 
\begin{equation*}
\int ds\;\bar{p}(s)=\frac{1}{\mathcal{V}},
\end{equation*}%
where $\mathcal{V}$ is a constant with units of a volume $[\Omega ]$. The
probability distribution $\bar{p}$ has units of $[\Omega ^{-1}V_{s}^{-1}]$,
where $V_{s}$ is the volume of the integration space of the parameters $s$.
Importantly, \textrm{Tr}$(\rho ^{\mathrm{imp}})=\nicefrac{\Omega}{%
\mathcal{V}}$ scales as $\Omega $, and so certainly $\rho ^{\mathrm{imp}%
}\neq \rho ^{\mathrm{th}}$, since \textrm{Tr}$(\rho ^{\mathrm{th}})=1$,
independent of the volume. \ Yet one might hope that the trace-improper
mixture could lead to a correct representation of \textit{some }of the
properties of thermal light, if $\bar{p}(s)$ and the spectral functions $f_{%
\mathbf{r}_{\mathrm{o}}s}$ and amplitudes $\alpha _{\mathbf{r}_{\mathrm{o}%
}s} $ are chosen correctly. \ We next show that this is possible for the
first-order correlation function (\ref{thermal1}). \ Here we let the volume $%
\Omega \rightarrow \infty $ in (\ref{rhoimp}) at the start, and show that
with a correct choice of $\bar{p}(s)$, $f_{\mathbf{r}_{\mathrm{o}}s}$, and $%
\alpha _{\mathbf{r}_{\mathrm{o}}s}$ we can find \ 
\begin{eqnarray*}
G_{ij}^{(1)\mathrm{imp}}(\mathbf{r}t;\mathbf{r}(t+\tau )) &\equiv \mathrm{Tr%
}\left( \rho ^{\mathrm{imp}}{E}_{i}^{(-)}(\mathbf{r},t){E}_{j}^{(+)}(\mathbf{%
r},t+\tau )\right) \\
=\int ds\int d\mathbf{r}_{\mathrm{o}}&\bar{p}(s) \:\left( \mathcal{E}_{i}^{(%
\mathbf{r}_{\mathrm{o}}s)}(\mathbf{r},t)\right) ^{\ast }\: \mathcal{E}_{j}^{(%
\mathbf{r}_{\mathrm{o}}s)}(\mathbf{r},t+\tau )\: ,
\end{eqnarray*}%
where the integral of $\mathbf{r}_{\mathrm{o}}$ now ranges over all space,
such that 
\begin{equation}
G_{ij}^{(1)\mathrm{imp}}(\mathbf{r}t;\mathbf{r}(t+\tau ))=G_{ij}^{(1)\mathrm{%
th}}(\mathbf{r}t;\mathbf{r}(t+\tau )).  \label{simulation}
\end{equation}

As a first example we take $s$ to include a central wave vector $\mathbf{k}_{%
\mathrm{o}}=k_{\mathrm{o}}\mathbf{\hat{m}}$ of the pulse, with the unit
vector $\mathbf{\hat{m}}$ identifying the main polarization direction, and a
unit vector $\mathbf{\hat{n}}$ that characterizes the polarization as
described below; thus $s=\{k_{\mathrm{o}},\mathbf{\hat{m}},\mathbf{\hat{n}}%
\} $, $V_{s}$ has units of inverse volume, and $\bar{p}(s)$ is
dimensionless. \ We characterize pulses of this type by 
\begin{equation*}
K(s\mathbf{,k}\lambda )=\mathcal{N\;}L(\mathbf{k,k}_{\mathrm{o}})\,\left( 
\mathbf{e}_{\mathbf{k\lambda }}^{\ast }\right) \cdot (\mathbf{k\times \hat{n}%
}),
\end{equation*}%
(recall Eq. \ref{Kdef}) where $\mathcal{N}$ is a normalization constant
chosen so the normalization of $f_{\mathbf{r}_{\mathrm{o}}s;\mathbf{k\lambda 
}}$ is satisfied, and $L(\mathbf{k,k}_{\mathrm{o}})$ is a real function. The
expectation value of $\mathbf{E}^{(+)}(\mathbf{r})\equiv \mathbf{E}^{(+)}(%
\mathbf{r},0)$ in the state $\left\vert \alpha _{\mathbf{r}_{\mathrm{o}}s}f_{%
\mathbf{r}_{\mathrm{o}}s}\right\rangle $ is given by \ 
\begin{equation*}
\mathcal{E}^{(\mathbf{r}_{\mathrm{o}}s)}(\mathbf{r})={}i\mathcal{N}\alpha _{%
\mathbf{r}_{\mathrm{o}}s}\int d\mathbf{k}\;\sqrt{\frac{\hbar \omega _{k}}{%
16\pi ^{3}\epsilon _{0}}}(\mathbf{k\times \hat{n}})L(\mathbf{k,k}_{\mathrm{o}%
})e^{i\mathbf{k\cdot (r-r}_{\mathrm{o}})}\,.
\end{equation*}%
This allows the representation of very general forms of pulses in free
space; $\mathcal{E}^{(\mathbf{r}_{\mathrm{o}}s)}(\mathbf{r})$ will typically
be centered at $\mathbf{r}_{\mathrm{o}}$, and its polarization is
characterized by having \textit{no }component in the $\mathbf{\hat{n}}$
direction, $\mathbf{\hat{n}}\cdot $ $\mathcal{E}^{(\mathbf{r}_{\mathrm{o}%
}s)}(\mathbf{r})=0$. \ In specifying $\bar{p}(s)$ we assume a uniform
distribution over pulse directions $\mathbf{\hat{m}}$; for a given $\mathbf{%
\hat{m}}$ only $\mathbf{\hat{n}}$ perpendicular to $\mathbf{\hat{m}}$ are
chosen, but the distribution over such $\mathbf{\hat{n}}$ is also uniform. \
Choosing a fixed direction in the plane perpendicular to $\mathbf{\hat{m}}$
for each $\mathbf{\hat{m}}$, and denoting the angle that $\mathbf{\hat{n}}$
makes from this direction by $\Psi $, we have 
\begin{equation*}
\int ds\;\bar{p}(s)\rightarrow \int_{0}^{\infty }dk_{\mathrm{o}}\;p(k_{%
\mathrm{o}})\int d\mathbf{\hat{m}}\int_{0}^{2\pi }d\Psi ,
\end{equation*}%
where the remaining dependence $p(k_{\mathrm{o}})$ is on the magnitude of
the central wave vector of the pulse, $k_{\mathrm{o}}=\left\vert \mathbf{k}_{%
\mathrm{o}}\right\vert $. \ 

For any proposed $L(\mathbf{k,k}_{\mathrm{o}})$ our task is then to see if $%
p(k_{\mathrm{o}})$ can be chosen so that our condition (\ref{simulation}) is
guaranteed. \ We begin by considering pulses of a Gaussian form; 
\begin{equation}
L(\mathbf{k,k}_{\mathrm{o}})=e^{-\frac{\left\vert \mathbf{k}-\mathbf{k}_{%
\mathrm{o}}\right\vert ^{2}}{2\sigma ^{2}}}.\label{Gaussian}
\end{equation}%
For thermal radiation at $T=5777$~K we find that the condition (\ref%
{simulation}) can be guaranteed only if $\sigma $ is chosen so that the
pulse has a bandwidth on the order of THz or smaller, describing pulses that
are on the order of picoseconds in length or longer \cite{Chenu2014c}. \
Interestingly, no physical solution can be found for femtosecond pulses with
a bandwidth as broad as the thermal spectrum. \ The problem is that the
Gaussian shape (\ref{Gaussian}) differs too much from the shape required to
guarantee that the norm of the integrand of (\ref{thermal1}) is reproduced.
\ Thus the only way that we can satisfy (\ref{simulation}) is to choose $%
\sigma $ so small that, compared with the thermal spectrum, $L(\mathbf{k,k}_{%
\mathrm{o}})$ is essentially proportional to a Dirac delta function; then $%
p(k_{\mathrm{o}})$ itself is relied on to capture the shape of that
integrand.

To satisfy (\ref{simulation}) with broadband pulses we can work instead with
a set of parameters $s$ that includes only a nominal direction of
propagation of the pulse $\mathbf{\hat{m}}$, as well as a polarization
vector $\mathbf{\hat{n}}$ as before, $s=\{\mathbf{\hat{m}},\mathbf{\hat{n}}%
\} $. Note now that $V_{s}$ is dimensionless whereas $\bar{p}(s)$ has
dimension of $\Omega ^{-1}$. \ We take our pulses (\ref{Kdef}) to be
specified by 
\begin{equation}
K(s\mathbf{,k}\lambda )=\mathcal{N}\;l(k)\;\upsilon (\mathbf{\hat{k}\cdot 
\hat{m})}\,\left( \mathbf{e}_{\mathbf{k\lambda }}^{\ast }\right) \cdot (%
\mathbf{k\times \hat{n}}),  \label{eq:thermal_def}
\end{equation}%
where the function $\upsilon (x)$ is chosen to characterize the spread in
the direction of wave vectors in the pulse and should be peaked at $x=1$ for 
$\mathbf{\hat{m}}$ to indicate the nominal direction of propagation of the
pulse; the function $l(k)$ is now relied on to help capture the shape of the
norm of the integrand of (\ref{thermal1}). For $\bar{p}(s)$ we assume that
the $\mathbf{\hat{m}}$ are distributed isotropically and, for each $\mathbf{%
\hat{m}}$, all $\mathbf{\hat{n}}$ perpendicular to $\mathbf{\hat{m}}$ are
equally distributed,%
\begin{equation*}
\int ds\;\bar{p}(s)\rightarrow p\int d\mathbf{\hat{m}}\int_{0}^{2\pi }d\Psi ,
\end{equation*}%
where $p$ is now a constant with units of $[\Omega ^{-1}]$, $d\mathbf{\hat{m}%
}$ indicates an integration over solid angle, and $\Psi $ denotes the angle $%
\mathbf{\hat{n}}$ makes from a fixed direction in the plane perpendicular to 
$\mathbf{\hat{m}}$. Such a trace-improper mixture can lead to (\ref%
{simulation}) by choosing 
\begin{equation}
l(k)=\frac{1}{k\sqrt{e^{\beta \hbar ck}-1}}\text{ and }p|\alpha |^{2}=\frac{%
4\zeta (3)}{\pi ^{4}(\beta c\hbar )^{3}}\,,  \label{lresult}
\end{equation}%
where $\zeta $ is the Riemann-zeta function \cite{Chenu2014c}. By comparing (%
\ref{thermal1}) and (\ref{lresult}), it is clear that the bandwidth of the
pulse helps capture that of the thermal radiation.

The requirement of a fixed product $p |\alpha|^2$ illustrates that the need
for improper behavior as $\Omega \to \infty$ can be met either by the trace
or by the pulses. In the example of a density operator $\rho^{\mathrm{imp}}$
we have been considering, any finite $p$ will lead to an infinite trace for $%
\Omega \to \infty$. Alternately, we could repeat the derivation sketched
here insisting on a density operator of unit trace; then we would find that $%
p$ would vanish as $\Omega\to\infty$, and the condition (\ref{lresult})
would demand that $|\alpha|^2$ diverge in that limit. The latter option
would only make physical sense for a finite volume of observation $\Omega$;
we put it aside for now, but return to it again below.

\begin{figure}[tbp]
\includegraphics[width=1\columnwidth]{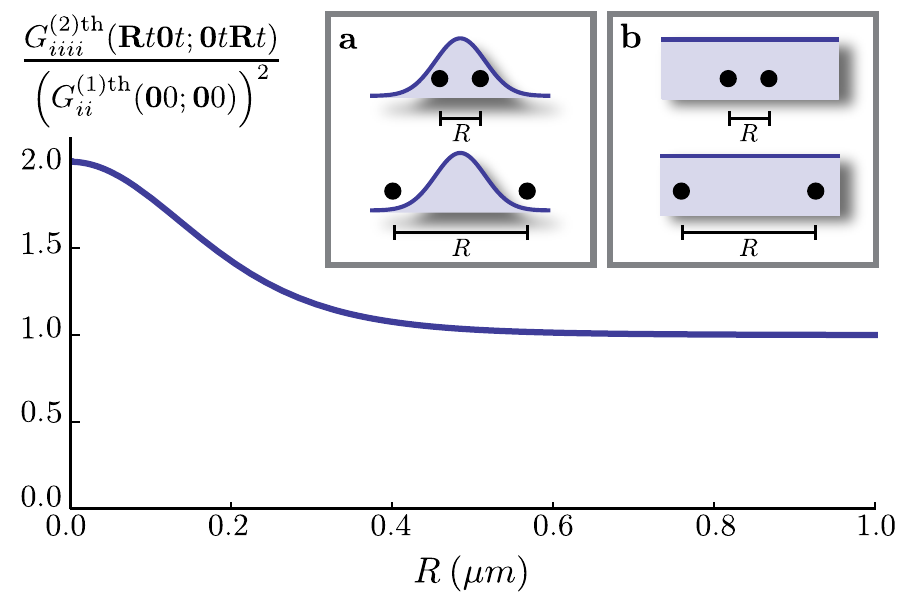} 
\caption{Normalized second-order correlation function for thermal light as a
function of the distance $R\mathbf{\hat{\imath}}$. Note that it is non-zero for all values of $R$%
. Inset: Schematic depicting two detectors (e.g. atoms with a broad
absorption band) within (a) the field of a localized pulse and (b) thermal
radiation. As the distance between the two atoms increases, the probability
that both absorb a photon from any localized pulse tends to zero. }
\label{fig:G2}
\end{figure}

The preceding two examples show that although a trace-improper mixture of
{\aurelia single} pulses is not described by the same density operator as thermal light, such
a mixture can be constructed to reproduce the first-order correlation
function of thermal light. \ As we look at higher-order correlation
functions we will necessarily find that such a mixture fails to reproduce
the properties of thermal light, since $\rho ^{\mathrm{imp}}\neq \rho ^{%
\mathrm{th}}$. \ But can such a mixture capture the second-order correlation
function? This is defined by ${G}_{i_{1}i_{2}i_{3}i_{4}}^{(2)}(\mathbf{r}_{1}%
{t}_{1}\mathbf{r}_{2}{t}_{2};\mathbf{r}_{3}{t}_{3}\mathbf{r}_{4}{t}%
_{4})=\langle E_{i_{1}}^{(-)}(\mathbf{r}_{1}{t}_{1})E_{i_{2}}^{(-)}(\mathbf{r%
}_{2}{t}_{2})E_{i_{3}}^{(+)}(\mathbf{r}_{3}{t}_{3})E_{i_{4}}^{(+)}(\mathbf{r}%
_{4}{t}_{4})\rangle $. Choosing all times identical along with $\mathbf{r}%
_{1}=\mathbf{r}_{4}$ and $\mathbf{r}_{2}=\mathbf{r}_{3}$, this expression
represents two simultaneous absorption events at positions $\mathbf{r}_{1}$
and $\mathbf{r}_{2}$ respectively \cite{Glauber1963a}. For thermal light at $%
T=5777$ K, the second-order correlation function $G_{iiii}^{(2)\mathrm{th}}(%
\mathbf{R}t\mathbf{0}t;\mathbf{0}t\mathbf{R}t)$ (no sum over $i$) is shown
in Fig. \ref{fig:G2} \cite{Branczyk2015a} as a function of the distance $%
\mathbf{R}\equiv \mathbf{r}_{2}-\mathbf{r}_{1}$ we imagine separating two
broadband detectors; we take the direction of $\mathbf{R}$ to lie along the
Cartesian axis $i$. Importantly, beyond a distance {\aurelia that corresponds to the 
coherence length, i.e. }of about {\aurelia $0.39~\mu m$} for $%
T=5777$ K, the second-order correlation function is independent of both the
distance $\mathbf{R}$ and the orientation of that vector with respect to the
Cartesian axis $i$, 
\begin{equation}
G_{iiii}^{(2)\mathrm{th}}(\mathbf{R}t\mathbf{0}t;\mathbf{0}t\mathbf{R}t)%
\underset{R\rightarrow \infty }{\longrightarrow }\left( \frac{\pi ^{2}}{%
90\epsilon _{0}\beta ^{4}(\hbar c)^{3}}\right) ^{2}.  \label{eq:G2_th_inf}
\end{equation}%
Clearly, in the presence of thermal light there is a non-zero probability of
simultaneous broadband detection events occurring regardless of the distance
between the two detectors.

\begin{figure*}[t]
\includegraphics[width=\textwidth]{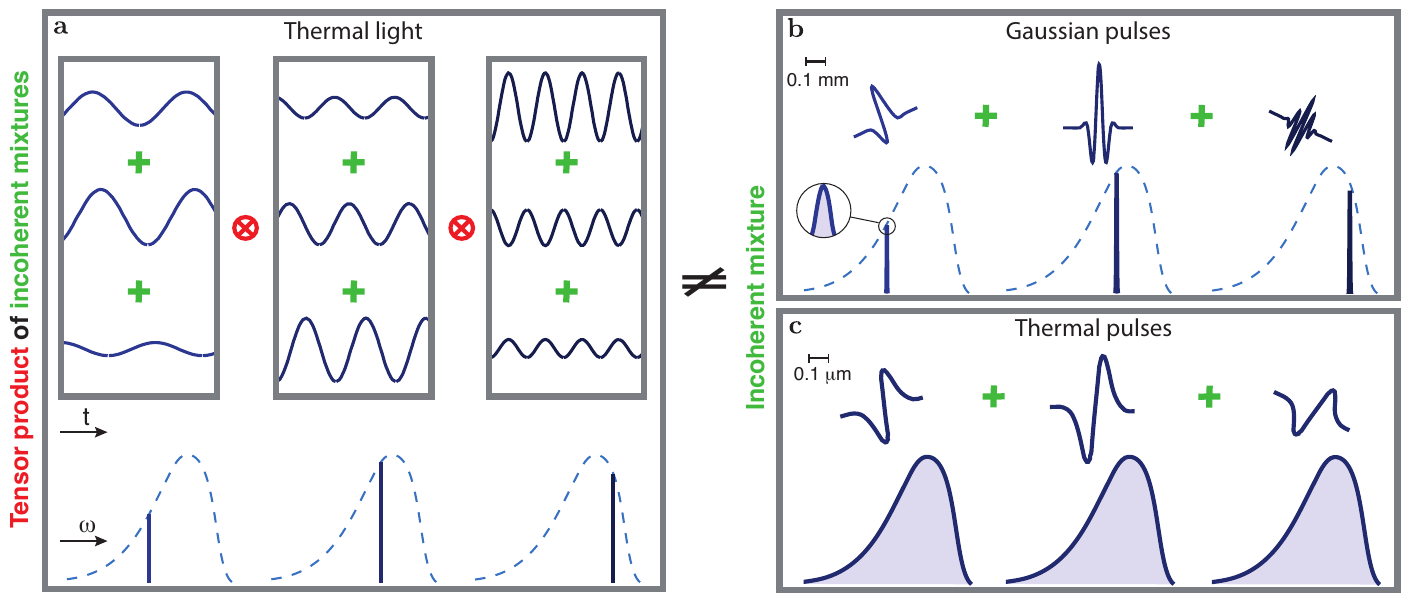}
\caption{Schematics illustrating the richness of thermal light as a tensor
product over an infinity of modes. The green ``$+$'' indicates an \emph{%
incoherent} sum; the red ``$\otimes$'' a direct product. (a) Thermal light
is built from a tensor product of mixed states, each composed by a mixture
of monochromatic coherent states (represented by continuous waves in the
insets) with various amplitudes and phases. (b;c) The trace-improper mixture
of {\aurelia single} pulses with (b) Gaussian or (c) thermal lineshape can only reproduce the
first-order correlation function (see details in the text). }
\label{fig:light_decomposition}
\end{figure*}

We have not been able to evaluate the second-order correlation function for $%
\rho ^{\mathrm{imp}}$ analytically. However, since each member of our family
of pulses is localized in space, each individual pulse is not able to
simultaneously excite two detectors if they are well separated. Although our
mixture is trace-improper, it is composed of an incoherent mixture of
{\aurelia individual,} localized pulses (\textit{cf.} Eq.~\ref{rhoimp}). Therefore the total
probability should equal the sum of the individual realizations, and we
expect our argument to apply to the mixture as well. Hence such a mixture
can never capture the result (\ref{eq:G2_th_inf}).

Returning to a finite observation volume $\Omega$ and the use of a
unit-trace density operator representing a mixture of {\aurelia single} pulses with $|\alpha_{%
\mathbf{r}_{\mathrm{o}} s}|^2$ proportional to $\Omega$, from the arguments
above we see that such a density operator would have no chance of describing
the second-order correlation function properly if the observation volume
were significantly larger than the size of the pulses. For if it were, the
chance of any pulse in the mixture exciting two detectors at different ends
of the observation volume would be negligible. Now at $T=5777$ K the size of
the pulses (\ref{eq:thermal_def}) is about 0.4 $\mu$m \cite{Chenu2014c}, as
might be expected from the characteristic length scale of Fig. 1. Hence even
if we employed a unit-trace mixture of {\aurelia single} pulses with the square of their
amplitudes proportional to the observation volume, for observation volumes
larger than a few cubic microns such mixtures would necessarily describe the
second-order correlation function incorrectly.

In summary, we have shown that no mixture of single coherent pulses can
represent thermal radiation. Allowing the mixture to be trace-improper, or
allowing the square of the amplitudes of the pulses to scale with the
observation volume, we can reproduce the first-order correlation function of
thermal light at equal space points. If Gaussian pulses are used, pulses
with a surprisingly narrow bandwidth are required. Alternately, broadband
pulses with a lineshape mirroring the thermal spectrum can be used. The
mixtures are schematically represented in Fig. \ref{fig:light_decomposition}%
. Nonetheless, these mixtures, and indeed any (proper or improper) mixture
of well-localized pulses, fail to reproduce even the second-order
correlation function of thermal light. The difficulties suggest that one
should look instead for a representation of thermal light as mixture of 
\emph{sets} of pulses, where each set contains more than one pulse. 

\begin{acknowledgements}
We are grateful to P. Brumer, B. Sanders, A. Steinberg and H. Wiseman for interesting and fruitful discussions.  We thank I. Kassal, S. Rahimi-Keshari and S. Baghbanzadeh for stressing the possibility of scaling the square of the pulse amplitudes with observation volume. A.C. acknowledges funding from the Swiss National Science Foundation and from the DFAIT of Canada.
 through the Post-Doctoral Research Fellowship awarded by the Government of Canada. 
This work was partly supported by the Natural Sciences and Engineering Research Council of Canada, DARPA (QuBE) and the United States Air Force Office of Scientific Research (FA9550-13-1-0005). Research at Perimeter Institute is supported by the Government of Canada through Industry Canada and by the Province of Ontario through the Ministry of Research and Innovation.
\end{acknowledgements}


\vspace*{-1ex}

\begin{thebibliography}{99}

\bibitem{Fleming1986a} G. Fleming, \textit{Chemical Applications of
Ultrafast Spectroscopy}, \textit{\ International Series of Monographs on
Chemistry} (Oxford University Press, 1986).

\bibitem{Zewail2000a} A.~H. Zewail, ``Femtochemistry: Atomic-Scale Dynamics
of the Chemical Bond,'' J. Phys. Chem. A \textbf{104}, 5660 (2000).

\bibitem{Grondelle2006a} R. van Grondelle and V.~I. Novoderezhkin, ``Energy
transfer in photosynthesis: experimental insights and quantitative models,''
Phys. Chem. Chem. Phys. \textbf{\ 8}, 793 (2006).

\bibitem{Cheng2009a} Y.-C. Cheng and G.~R. Fleming, ``Dynamics of Light
Harvesting in Photosynthesis,'' Annu. Rev. Phys. Chem. \textbf{60}, 241
(2009).

\bibitem{Jiang1991a} X.-P. Jiang and P. Brumer, ``Creation and dynamics of
molecular states prepared with coherent vs partially coherent pulsed
light,'' J. Chem. Phys. \textbf{94}, 5833 (1991).

\bibitem{Mancal2010a} T. Man\v{c}al and L. Valkunas, ``Exciton dynamics in
photosynthetic complexes: excitation by coherent and incoherent light,'' New
Journal of Physics \textbf{\ 12}, 065044 (2010).

\bibitem{Hoki2011a} K. Hoki and P. Brumer, ``Excitation of biomolecules by
coherent vs. incoherent light: Model rhodopsin photoisomerization,'' in 
\textit{Procedia Chemistry}, \textbf{3}, 122 (2011).


\bibitem{Brumer2012a} P. Brumer and M. Shapiro, ``Molecular response in
one-photon absorption via natural thermal light vs. pulsed laser
excitation,'' Proc. Nat. Am. Soc. \textbf{\ 109}, 19575 (2012).

\bibitem{Kassal2013a} I. Kassal, J. Yuen-Zhou, and S. Rahimi-Keshari, ``Does
Coherence Enhance Transport in Photosynthesis?,'' J. Phys. Chem. L. \textbf{4%
}, 362 (2013).

\bibitem{Kano1962a} Y. Kano and E. Wolf, ``Temporal coherence of black body
radiation,'' Proc. Phys. Soc. \textbf{80}, 1273 (1962).

\bibitem{Iqbal1983a} M. Iqbal, \textit{An Introduction to Solar Radiation}
(Academic Press, 1983).

\bibitem{MandelWolfBook} L. Mandel and E. Wolf, \textit{Optical coherence
and quantum Optics} (Cambridge university press, 1995), {C}hap. 13.

\bibitem{Glauber1963a} R.~J. Glauber, ``Coherent and incoherent states of
the radiation field,'' Phys. Rev. \textbf{131}, 2766 (1963).

\bibitem{Loudon2000} R. Loudon, \textit{The Quantum Theory of Light}, 3rd
ed. (Oxford University Press, Oxford, 2000), {C}hap. 1.

\bibitem{Chenu2014c} A. Chenu, A.~M. Bra\'nczyk, and J.~E. Sipe,
``First-order decomposition of thermal light in terms of a statistical
mixture of pulses,'' \textit{in preparation}.

\bibitem{Bradley1974a} D.~G. Bradley and G.~H.~C. New, ``Ultrashort pulse
measurement,'' in \textit{\ IEEE}, \textbf{72}, 313 (1974).

\bibitem{Mehta1964b} C.~L. Mehta and E. Wolf, ``Coherence Properties of
Blackbody Radiation. {II}. {C}orrelation tensors of the quantized field,''
Phys. Rev. \textbf{134}, 1149 (1964).

\bibitem{Branczyk2015a} A.~M. Bra\'nczyk, A. Chenu, and J.~E. Sipe,
``Second-order correlation function of thermal light,'' \textit{in
preparation}.

\end{thebibliography}
\end{document}